\documentclass[a4paper,12pt]{article}

\usepackage[english]{babel}
\usepackage[dvips]{graphicx}

\usepackage{cite}

\usepackage{latexsym,amssymb, amsthm}
\usepackage[centertags]{amsmath}
\usepackage{epsfig,pifont,rotating}
\usepackage[inactive]{srcltx}  

\usepackage{helvet,times,mathptm,epic,eepic}

\usepackage{color}

\usepackage{newlfont}

\usepackage{amsfonts}

\usepackage{times}
\def\beq{\begin{equation}}
\def\eeq{\end{equation}}

\title{Interpretation of percolation in terms of infinity computations}
\author{D.I. Iudin\footnote{Dmitry I. Iudin, Ph.D., D.Sc., holds a Full Professorship
at the N.I.~Lobatchevsky State University,
Nizhni Novgorod, Russia. He is also Full Professor   at the Architecture State University,
Nizhni Novgorod, Russia and Senior Researcher at the Institute
of Applied physics of Russian Academy of Science, Nizhni Novgorod, Russia.
{\tt  di-iudin@nirfi.sci-nnov.ru}},
 Ya.D. Sergeyev\footnote{Yaroslav D. Sergeyev, Ph.D., D.Sc., holds a Full Professorship
 reserved for distinguished scientists at the University of Calabria, Rende, Italy.
 He is also Full Professor   at the N.I.~Lobatchevsky State University,
Nizhni Novgorod, Russia and Affiliated Researcher at the Institute
of High Performance   Computing and Networking of the National
Research Council of Italy. {\tt  yaro@si.deis.unical.it}}, M.
Hayakawa\footnote{Masaschi Hayakawa, Ph.D., D.Sc., holds a Full Professorship
at the Department of Electronic Engineering, University of Electro-Communications,
Chofu, Tokyo 182 8585, Japan. {\tt  hayakawa@whistler.ee.uec.ac.jp}}}
\date{}

\begin{document}
\maketitle

\begin{center}
\textbf{Abstract}
\end{center}

In this paper, a number of traditional models related to the
percolation theory has been considered by means of new computational
methodology that does not use Cantor's ideas and describes
infinite and infinitesimal numbers in accordance with the
principle `The part is less than the whole'. It  gives a
possibility to work with finite, infinite, and infinitesimal
quantities \textit{numerically} by using a new kind of a computer
-- the Infinity Computer -- introduced recently in
 \cite{Sergeyev_patent}.  The new approach   does not contradict Cantor. In contrast,
it can be viewed as an evolution of his deep ideas regarding the
existence of different infinite numbers in a more applied way.
Site percolation and gradient percolation have been studied by
applying the new computational tools. It has been established that
in an infinite system the phase transition point is not really a
point as with respect of traditional approach. In light of new
arithmetic it appears as a critical interval, rather than a
critical point. Depending on ``microscope" we use this interval
could be regarded as finite, infinite and infinitesimal short
interval. Using new approach we observed that in vicinity of
percolation threshold we have {\it many} different {\it infinite
clusters} instead of {\it one infinite cluster} that appears in
traditional consideration.

\newpage

\section{Introduction}
\label{s1}

 Numerous trials
have been done during the centuries in order to evolve existing
numeral systems\footnote{ We are reminded that a \textit{numeral}
is a symbol or group of symbols that represents a \textit{number}.
The difference between numerals and numbers is the same as the
difference between words and the things they refer to. A
\textit{number} is a concept that a \textit{numeral} expresses.
The same number can be represented by different numerals. For
example, the symbols `8', `eight', and `VIII' are different
numerals, but they all represent the same number.} in such a way
that infinite and infinitesimal numbers could be included in them
(see \cite{Benci,Cantor,Conway,Leibniz,Newton,Robinson,Wallis}).
Particularly, in the early history of the calculus, arguments
involving infinitesimals played a pivotal role in the derivation
developed by Leibnitz and Newton (see \cite{Leibniz,Newton}). The
notion of an infinitesimal, however, lacked a precise mathematical
definition and in order  to provide a more rigorous   foundation
for the calculus infinitesimals were gradually replaced by the
d'Alembert-Cauchy  concept of a limit (see
\cite{Cauchy,DAlembert}).

The creation of a mathematical theory of infinitesimals to base on the calculus remained an open problem until the end of
1950s when Robinson (see \cite{Robinson})   introduced his famous
non-standard analysis approach. He showed that non-archimedean
ordered field extensions of the reals contained numbers that could
serve the role of infinitesimals and their reciprocals   could
serve as infinitely large numbers. Robinson then has derived the
theory of limits,  and more generally of Calculus, and has found a
number of important applications of his ideas in many other fields
of Mathematics (see \cite{Robinson}).

In his approach,  Robinson used mathematical tools and terminology
(cardinal numbers, countable sets, continuum, one-to-one
correspondence, etc.) taking their origins from the famous ideas
of Cantor (see \cite{Cantor}) who has shown that there existed
infinite sets having different number of elements. It is well
known nowadays that while dealing with infinite sets, Cantor's
approach leads to some counterintuitive situations that often are
called by non-mathematicians `paradoxes'. For example, the set of
even numbers, $\mathbb{E}$, can be put in a one-to-one
correspondence with the set of all natural numbers, $\mathbb{N}$,
in spite of the fact that $\mathbb{E}$ is  a part of $\mathbb{N}$:
 \beq
\begin{array}{lccccccc}
  \mbox{even numbers:}   & \hspace{5mm} 2, & 4, & 6, & 8,  & 10, & 12, &
\ldots    \\

& \hspace{5mm} \updownarrow &  \updownarrow & \updownarrow  &
\updownarrow  & \updownarrow  &  \updownarrow &   \\

  \mbox{natural numbers:}& \hspace{5mm}1, &  2, & 3, & 4 & 5,
       & 6,  &    \ldots \\
     \end{array}
\label{4.4.1}
 \eeq
The philosophical principle of Ancient Greeks `\textit{The part is
less than the whole}' observed in the world around us does not
hold true for infinite   numbers   introduced by Cantor, e.g., it
follows $x+1=x$, if $x$ is an infinite cardinal,  although for any
finite $x$ we have   $x+1>x$. As a consequence, the same effects
necessary have reflections in the non-standard Analysis of
Robinson (this is not the case of the interesting non-standard
approach introduced recently in \cite{Benci}).

Due to the enormous importance of the concepts of infinite and
infinitesimal in science, people try to introduce them in their
work with computers, too  (see, e.g. the IEEE Standard for Binary
Floating-Point Arithmetic). However, non-standard Analysis remains
a very theoretical field   because various arithmetics (see
\cite{Benci,Cantor,Conway,Robinson}) developed for infinite and
infinitesimal numbers are quite different with respect to the
finite arithmetic we are used to deal with. Very often they leave
undetermined many operations where infinite numbers take part (for
example, $\infty-\infty$, $\frac{\infty}{\infty}$,  sum of
infinitely many items, etc.) or use representation of infinite
numbers based on infinite sequences of finite numbers. These
crucial difficulties did  not allow people to construct computers
that would be able to work with infinite and infinitesimal numbers
\textit{in the same manner} as we are used to do with finite
numbers  and to study infinite and infinitesimal objects
numerically.

Recently a new applied point of view on infinite and infinitesimal
numbers  has been introduced in
\cite{Sergeyev,informatica,Lagrange}. The new approach does not
use Cantor's ideas and describes infinite and infinitesimal
numbers that are in accordance with the principle `The part is
less than the whole'. It  gives a possibility to work with finite,
infinite, and infinitesimal quantities \textit{numerically} by
using a new kind of computers -- the Infinity Computer --
introduced in \cite{Sergeyev_patent,www,Sergeyev11a,Sergeyev11b}.  It is worthwhile
noticing that the new approach   does not contradict Cantor. In
contrast, it can be viewed as an evolution of his deep ideas
regarding the existence of different infinite numbers in a more
applied way. For instance, Cantor showed that there exist infinite
sets having different cardinalities $\aleph_0$ and $\aleph_1$. In
its turn, the new approach specifies this result showing that in
certain cases within each of these classes it is possible to
distinguish sets with the number of elements being different
infinite numbers. We   emphasize   that the new approach has been
introduced as an evolution of standard and non-standard Analysis
and not as a contraposition to them. One or another version of
Analysis can be chosen by the working mathematician in dependence
on the problem he deals with.


In this paper, we consider a number of  applications related to
the theory of percolation and study them using the new approach.
On the one hand, percolation represents the simplest model of a
disordered system. Disordered structures and random processes that
are self-similar on certain length and time scales are very common
in nature. They can be found on the largest and the smallest
scales: in galaxies and landscapes, in earthquakes and fractures,
in aggregates and colloids, in rough surfaces and interfaces, in
glasses and polymers, in proteins and other large molecules.
Disorder plays a fundamental role in many processes of industrial
and scientific interest. On the other hand, percolation reveals a
concept of self-similarity and demonstrate numerous fractal
features. Owing to the wide occurrence of self-similarity in
nature, the scientific community interested in this phenomenon is
very broad, ranging from astronomers and geoscientists to material
scientists and life scientists. From mathematic point of view,
self-similarity implies a recursive process and, consequently, is
tightly connected with concept of infinity. This turn us to an
idea that percolation is very suitable to demonstrate advantages
of the new computational approach  proposed recently in
\cite{Sergeyev,informatica}.

The outline of the paper is as follows. In Sec.~\ref{s2} we introduce the
new approach that allows one to write
down different finite, infinite, and infinitesimal numbers by a
finite number of symbols as particular cases of a unique framework
and to execute numerical computations with all of them. Than in Sec.~\ref{s3}
we apply the new methodology to the percolation phase transition.
Generalized percolation problem known as gradient percolation analyzed in terms of infinity computations in Sec.~\ref{s4}
In the final section, the applications are summarized and discussed.

\section{Methodology}
\label{s2}

 In this section, we give   a brief
introduction to the new methodology    that can be found in a
rather comprehensive form in   \cite{informatica,Lagrange}
downloadable from \cite{www} (see also the monograph
\cite{Sergeyev} written in a popular manner). A number of
applications of the new approach can be found in
\cite{chaos,spirals,icomp,Korea,Margenstern,Sergeyev11a,Sergeyev11b}. We start by introducing three
postulates that will fix our methodological positions (having a
strong applied character) with respect to infinite and
infinitesimal quantities and Mathematics, in general.

Usually, when mathematicians deal with infinite objects (sets or
processes) it is supposed   that human beings are able to execute
certain operations infinitely many times (e.g., see
(\ref{4.4.1})). Since we live in a finite world and all human
beings and/or computers finish operations they have started, this
supposition is not adopted.

 \textbf{Postulate 1.} \textit{There  exist
infinite and infinitesimal objects but   human beings and machines
are able to execute only a finite number of operations.}

Due to this Postulate, we accept a priori that we shall never be
able to give a complete description of infinite processes and sets
due to our finite capabilities.

The second postulate is adopted  following the way of reasoning
used in natural sciences where researchers use tools to describe
the object of their study and the instrument used influences the
results of observations. When physicists see a black dot in their
microscope they cannot say: The object of observation \textit{is}
the black dot. They are obliged to say: the lens used in the
microscope allows us to see the black dot and it is not possible
to say anything more about the nature of the object of observation
until we  change the instrument - the lens or the microscope
itself - by a more precise one.

Due to Postulate 1, the same happens in Mathematics studying
natural phenomena, numbers, and objects that can be constructed by
using numbers. Numeral systems used to express numbers are among
the instruments of observations used by mathematicians. Usage of
powerful numeral systems gives the possibility to obtain more
precise results in mathematics and in the same way usage of a good
microscope gives the possibility of obtaining more precise results
in Physics. However, the capabilities of the tools will be always
so limited due to Postulate 1 and due to Postulate~2 that we shall never
tell, \textbf{what is}, for example, a number but shall just
observe it through numerals expressible in a chosen numeral
system.

 \textbf{Postulate
2.} \textit{We shall not   tell \textbf{what are} the mathematical
objects we deal with; we just shall construct more powerful tools
that will allow us to improve our capacities to observe and to
describe properties of mathematical objects.}

Particularly, this means that from our point of view, axiomatic
systems do not define mathematical objects but just determine
formal rules for operating with certain numerals reflecting some
properties of the studied mathematical objects. Throughout the
paper, we shall always emphasize this philosophical triad --
researcher, object of investigation, and tools used to observe the
object -- in various mathematical and computational contexts.

Finally, we adopt the principle of Ancient Greeks mentioned above
as  the third postulate.

\textbf{Postulate 3.} \textit{The principle `The part is less than
the whole' is applied to all numbers (finite, infinite, and
infinitesimal) and to all sets and processes (finite and
infinite).}

Due to this declared applied statement, it becomes clear that the
subject of this paper is out of Cantor's approach and, as a
consequence, out of non-standard analysis of Robinson. Such
concepts as bijection, numerable and continuum sets, cardinal and
ordinal numbers cannot be used in this paper because they belong
to the theory working with different assumptions. However, the
approach used here does not contradict Cantor and Robinson. It can
be viewed just as a more strong lens of a mathematical microscope
that allows one to distinguish more objects and to work with them.

In \cite{Sergeyev,informatica}, a
  new  numeral system has
been developed in accordance with Postulates 1--3. It gives one a
possibility to execute numerical computations not only with finite
numbers but also with infinite and infinitesimal ones. The main
idea consists of the possibility to measure infinite and
infinitesimal quantities by different (infinite, finite, and
infinitesimal) units of measure.

A new infinite unit of measure   has been introduced for this
purpose   as the number of elements of the set $\mathbb{N}$ of
natural numbers. It is expressed by the numeral \ding{172} called
\textit{grossone}. It is necessary to note immediately that
\ding{172} is neither Cantor's $\aleph_0$ nor $\omega$.
Particularly, it has both cardinal and ordinal properties as usual
finite natural numbers (see \cite{informatica}).

Formally, grossone is introduced as a new number by describing its
properties postulated by the \textit{Infinite Unit Axiom} (IUA)
(see \cite{Sergeyev,informatica}). This axiom is added to axioms
for real numbers similarly to addition of the axiom determining
zero to axioms of natural numbers when integer numbers are
introduced. It is important to emphasize that we speak about
axioms of real numbers in sense of Postulate~2, i.e., axioms
define formal rules of operations with numerals in a given numeral
system.

Inasmuch as it has been postulated that grossone is a number,  all
other axioms for numbers hold for it, too. Particularly,
associative and commutative properties of multiplication and
addition, distributive property of multiplication over addition,
existence of   inverse  elements with respect to addition and
multiplication hold for grossone as for finite numbers. This means
that  the following relations hold for grossone, as for any other
number
 \beq
 0 \cdot \mbox{\ding{172}} =
\mbox{\ding{172}} \cdot 0 = 0, \hspace{3mm}
\mbox{\ding{172}}-\mbox{\ding{172}}= 0,\hspace{3mm}
\frac{\mbox{\ding{172}}}{\mbox{\ding{172}}}=1, \hspace{3mm}
\mbox{\ding{172}}^0=1, \hspace{3mm}
1^{\mbox{\tiny{\ding{172}}}}=1, \hspace{3mm}
0^{\mbox{\tiny{\ding{172}}}}=0.
 \label{3.2.1}
       \eeq

Let us comment upon the nature of grossone by some illustrative
examples.

Infinite numbers constructed using grossone  can be
interpreted in terms of the number of elements of infinite sets.
For example, $\mbox{\ding{172}}-1$ is the number of elements of a
set $B=\mathbb{N}\backslash\{b\}$, $b \in \mathbb{N}$, and
$\mbox{\ding{172}}+1$ is the number of elements of a set
$A=\mathbb{N}\cup\{a\}$, where $a \notin \mathbb{N}$.  Due to
Postulate~3, integer positive numbers that are larger than
grossone do not belong to $\mathbb{N}$ but also can be easily
interpreted. For instance,   $\mbox{\ding{172}}^2$ is the number
of elements of the set
  $V$, where  $ V  =
\{ (a_1, a_2)  : a_1 \in   \mathbb{N}, a_2 \in   \mathbb{N}  \}.
 $   \hfill $\Box$


Grossone has been introduced as the quantity of
natural numbers. As a consequence, similarly  to the set
 \beq
  A=\{1, 2, 3, 4, 5\}
\label{4.1.deriva_0}
 \eeq
   consisting of
5 natural numbers where 5 is the largest number in $A$, \ding{172}
is the largest    number\footnote{This fact is one of the
important methodological differences with respect to non-standard
analysis theories where it is supposed that infinite numbers   do
not belong to $\mathbb{N}$.} in $\mathbb{N}$ and
$\mbox{\ding{172}} \in \mathbb{N}$ analogously to the fact that 5
belongs to $A$. Thus, the set, $\mathbb{N}$, of natural numbers
can be written  in the form
 \beq
\mathbb{N} = \{ 1,2,  \hspace{3mm} \ldots  \hspace{3mm}
\frac{\mbox{\ding{172}}}{2}-2, \frac{\mbox{\ding{172}}}{2}-1,
\frac{\mbox{\ding{172}}}{2}, \frac{\mbox{\ding{172}}}{2}+1,
\frac{\mbox{\ding{172}}}{2}+2, \hspace{3mm}  \ldots \hspace{3mm}
\mbox{\ding{172}}-2, \hspace{2mm}\mbox{\ding{172}}-1, \hspace{2mm}
\mbox{\ding{172}} \}.   \label{4.1}
       \eeq
Note that traditional numeral systems did not allow us to see
infinite natural numbers
 \beq \ldots  \hspace{3mm}
\frac{\mbox{\ding{172}}}{2}-2, \frac{\mbox{\ding{172}}}{2}-1,
\frac{\mbox{\ding{172}}}{2}, \frac{\mbox{\ding{172}}}{2}+1,
\frac{\mbox{\ding{172}}}{2}+2, \hspace{3mm} \ldots  \hspace{3mm}
\mbox{\ding{172}}-2, \mbox{\ding{172}}-1, \mbox{\ding{172}}.
\label{4.1.deriva_1}
 \eeq
Similarly,
  Pirah\~{a}\footnote{Pirah\~{a} is a primitive tribe living in
Amazonia  that uses a very simple numeral system for counting:
one, two, `many'(see \cite{Gordon}). For Pirah\~{a}, all
quantities larger than two are just `many' and such operations as
2+2 and 2+1 give the same result, i.e., `many'. Using their weak
numeral system Pirah\~{a} are not able to distinguish numbers
larger than 2 and, as a result, to execute arithmetical operations
with them. Another peculiarity of this numeral system  is that
\mbox{`many'}+ 1= \mbox{`many'}. It can be immediately seen that
this result  is very similar to our traditional record $\infty +
1= \infty$.}     are not able to see  finite numbers larger than 2
using their weak numeral system but these numbers are visible if
one uses a more powerful numeral system. Due to Postulate~2, the
same object  of observation -- the set $\mathbb{N}$ --   can be
observed by different instruments -- numeral systems -- with
different accuracies allowing one to express  more or less natural
numbers. \hfill $\Box$

This example illustrates also the fact that when we speak about
sets (finite or infinite) it is necessary to take care about tools
used to describe a set (remember Postulate~2). In order to
introduce a set, it is necessary to have a language (e.g., a
numeral system) allowing us to describe its elements and the
number of the elements in the set. For instance, the set $A$ from
(\ref{4.1.deriva_0}) cannot be defined using the mathematical
language of Pirah\~{a}.

Analogously, the words `the set of all finite numbers' do not
define a set completely from our point of view, as well. It is
always necessary to specify which instruments are used to describe
(and to observe) the required set and, as a consequence, to speak
about `the set of all finite numbers expressible in a fixed
numeral system'. For instance, for Pirah\~{a} `the set of all
finite numbers'  is the set $\{1, 2 \}$ and for another Amazonian
tribe -- Munduruk\'u\footnote{Munduruk\'u (see \cite{Pica}) fail
in exact arithmetic with numbers larger than   5 but are able to
compare and add large approximate numbers that are far beyond
their naming range. Particularly, they use the words `some, not
many' and `many, really many' to distinguish two types of large
numbers (in this connection think about Cantor's  $\aleph_0$ and
$\aleph_1$).} -- `the set of all finite numbers' is the set $A$
from (\ref{4.1.deriva_0}). As it happens in Physics, the
instrument used for an observation bounds the possibility of
observation. It is not possible to say how we shall see the object
of our observation if we have not clarified which instruments will
be used to execute the observation.

Introduction of  grossone gives us a possibility to compose new
(in comparison with traditional numeral systems) numerals and to
see through them not only numbers (\ref{4.1.deriva_0}) but also
certain numbers larger than \ding{172}. We can speak about the set
of \textit{extended natural numbers} (including $\mathbb{N}$ as a
proper subset) indicated as $\widehat{\mathbb{N}}$ where
 \beq
  \widehat{\mathbb{N}} = \{
1,2, \ldots ,\mbox{\ding{172}}-1, \mbox{\ding{172}},
\mbox{\ding{172}}+1, \mbox{\ding{172}}+2, \mbox{\ding{172}}+3,
\ldots , \mbox{\ding{172}}^2-1, \mbox{\ding{172}}^2.
\mbox{\ding{172}}^2+1, \ldots \} \label{4.2.2}
       \eeq
However, analogously to the situation with `the set of all finite
numbers', the number of elements of the set $\widehat{\mathbb{N}}$
cannot be expressed within a numeral system using only \ding{172}.
It is necessary to introduce in a reasonable way a more powerful
numeral system and to define   new numerals (for instance,
\ding{173}, \ding{174}, etc.) of this system that would allow one
to fix the set (or sets) somehow. In general, due to Postulate~1
and~2, for any fixed numeral $\mathcal{A}$ system there always be
sets that cannot be described using $\mathcal{A}$.


Analogously to (\ref{4.1}), the set, $\mathbb{E}$, of even natural
numbers can be written now in the form
 \beq
\mathbb{E} = \{ 2,4,6 \hspace{5mm} \ldots  \hspace{5mm}
\mbox{\ding{172}}-4, \hspace{2mm}\mbox{\ding{172}}-2, \hspace{2mm}
\mbox{\ding{172}} \}.   \label{4.1.0}
       \eeq
Due to Postulate 3 and the IUA (see \cite{Sergeyev,informatica}),
it follows that the number of elements of the set of even numbers
is equal to $\frac{\mbox{\ding{172}}}{2}$ and \ding{172} is even.
Note that the next even number is $\mbox{\ding{172}}+2$ but it is
not natural because $\mbox{\ding{172}}+2  > \mbox{\ding{172}}$, it
is extended natural (see (\ref{4.2.2})). Thus, we can write down
not only initial (as it is  done traditionally) but also the final
part of (\ref{4.4.1})
  \[
\begin{array}{cccccccccc}
 2, & 4, & 6, & 8,  & 10, & 12, & \ldots  &
\mbox{\ding{172}} -4,  &    \mbox{\ding{172}}  -2,   &
\mbox{\ding{172}}    \\
 \updownarrow &  \updownarrow & \updownarrow  &
\updownarrow  & \updownarrow  &  \updownarrow  & &
  \updownarrow    & \updownarrow   &
  \updownarrow
   \\
 1, &  2, & 3, & 4 & 5, & 6,   &   \ldots  &
\frac{\mbox{\ding{172}}}{2} - 2,  &
     \frac{\mbox{\ding{172}}}{2} - 1,  &    \frac{\mbox{\ding{172}}}{2}   \\
     \end{array}
\]
concluding so (\ref{4.4.1})   in a complete accordance with
Postulate~3. It is worth  noticing that the new numeral system
allows us to solve many other `paradoxes' related to infinite and
infinitesimal quantities (see \cite{Sergeyev,informatica,Korea}).
 \hfill $\Box$

In order to  express numbers having finite, infinite, and
infinitesimal parts, records similar to traditional positional
numeral systems can be used (see \cite{Sergeyev,informatica}). To
construct a number $C$ in the new numeral positional system with
base \ding{172}, we subdivide $C$ into groups corresponding to
powers of \ding{172}:
 \beq
  C = c_{p_{m}}
\mbox{\ding{172}}^{p_{m}} +  \ldots + c_{p_{1}}
\mbox{\ding{172}}^{p_{1}} +c_{p_{0}} \mbox{\ding{172}}^{p_{0}} +
c_{p_{-1}} \mbox{\ding{172}}^{p_{-1}}   + \ldots   + c_{p_{-k}}
 \mbox{\ding{172}}^{p_{-k}}.
\label{3.12}
       \eeq
 Then, the record
 \beq
  C = c_{p_{m}}
\mbox{\ding{172}}^{p_{m}}    \ldots   c_{p_{1}}
\mbox{\ding{172}}^{p_{1}} c_{p_{0}} \mbox{\ding{172}}^{p_{0}}
c_{p_{-1}} \mbox{\ding{172}}^{p_{-1}}     \ldots c_{p_{-k}}
 \mbox{\ding{172}}^{p_{-k}}
 \label{3.13}
       \eeq
represents  the number $C$, where all numerals $c_i\neq0$, they
belong to a traditional numeral system and are called
\textit{grossdigits}. They express finite positive or negative
numbers and show how many corresponding units
$\mbox{\ding{172}}^{p_{i}}$ should be added or subtracted in order
to form the number $C$.

Numbers $p_i$ in (\ref{3.13}) are  sorted in the decreasing order
with $ p_0=0$
\[
p_{m} >  p_{m-1}  > \ldots    > p_{1} > p_0 > p_{-1}  > \ldots
p_{-(k-1)}  >   p_{-k}.
 \]
They are called \textit{grosspowers} and they themselves can be
written in the form (\ref{3.13}).
 In the record (\ref{3.13}), we write
$\mbox{\ding{172}}^{p_{i}}$ explicitly because in the new numeral
positional system  the number   $i$ in general is not equal to the
grosspower $p_{i}$. This gives the possibility to write down
numerals without indicating grossdigits equal to zero.

The term having $p_0=0$ represents the finite part of $C$ because,
due to (\ref{3.2.1}), we have $c_0 \mbox{\ding{172}}^0=c_0$. The
terms having finite positive gross\-powers represent the simplest
infinite parts of $C$. Analogously, terms   having   negative
finite grosspowers represent the simplest infinitesimal parts of
$C$. For instance, the  number
$\mbox{\ding{172}}^{-1}=\frac{1}{\mbox{\ding{172}}}$ is
infinitesimal. It is the inverse element with respect to
multiplication for \ding{172}:
 \beq
\mbox{\ding{172}}^{-1}\cdot\mbox{\ding{172}}=\mbox{\ding{172}}\cdot\mbox{\ding{172}}^{-1}=1.
 \label{3.15.1}
       \eeq
Note that all infinitesimals are not equal to zero. Particularly,
$\frac{1}{\mbox{\ding{172}}}>0$ because it is a result of division
of two positive numbers. All of the numbers introduced above can
be grosspowers, as well, giving thus a possibility to have various
combinations of quantities and to construct  terms having a more
complex structure.

\section{Geometric phase transition}
\label{s3}

In 1957, two mathematicians, S.R. Broadbent and J.M. Hammersley,
have published an article \cite {Broadbent} where they have shared
with readers an idea of probabilistic formalizations of water
infiltration in electric coffee maker. Their description, named
later   \textit{percolation theory}, represents one of the
simplest models of a disordered system.

\begin{figure}[h!]
\centering\noindent\vspace{1cm}
\includegraphics[width=.75\textwidth,height=12cm]{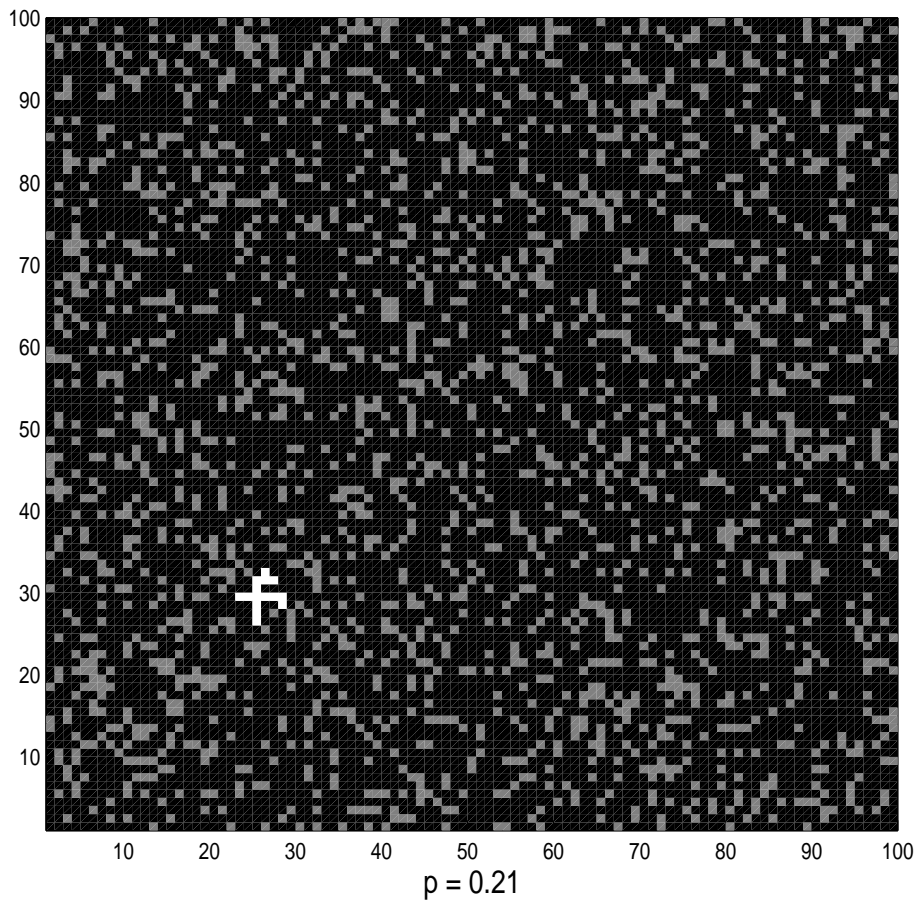}
\caption{Site percolation on a square lattice. Grey cells of a square lattice correspond to conducting pixels, black stand for non-conducting, white cells belong to maximal conducting cluster. Concentration of conducting pixels equals to $p=0,21 $}\label{f:pint1}
\end{figure}

Consider a square lattice, where each site is occupied randomly
with probability $p$ or empty with probability $1-p$. Occupied and
empty sites may stand for very different physical properties
\cite{BundeHav1,BundeHav,Feder,Stauff}. For simplicity, let us assume
that the occupied sites are electrical conductors (represented by
gray pixels in figure~\ref{f:pint1}), the empty sites (shown by
black pixels in figure~\ref{f:pint1}) represent insulators, and
that electrical current can flow only between nearest neighbor
conductor sites.

\begin{figure}[h!]
\centering\noindent\vspace{1cm}
\includegraphics[width=\textwidth]{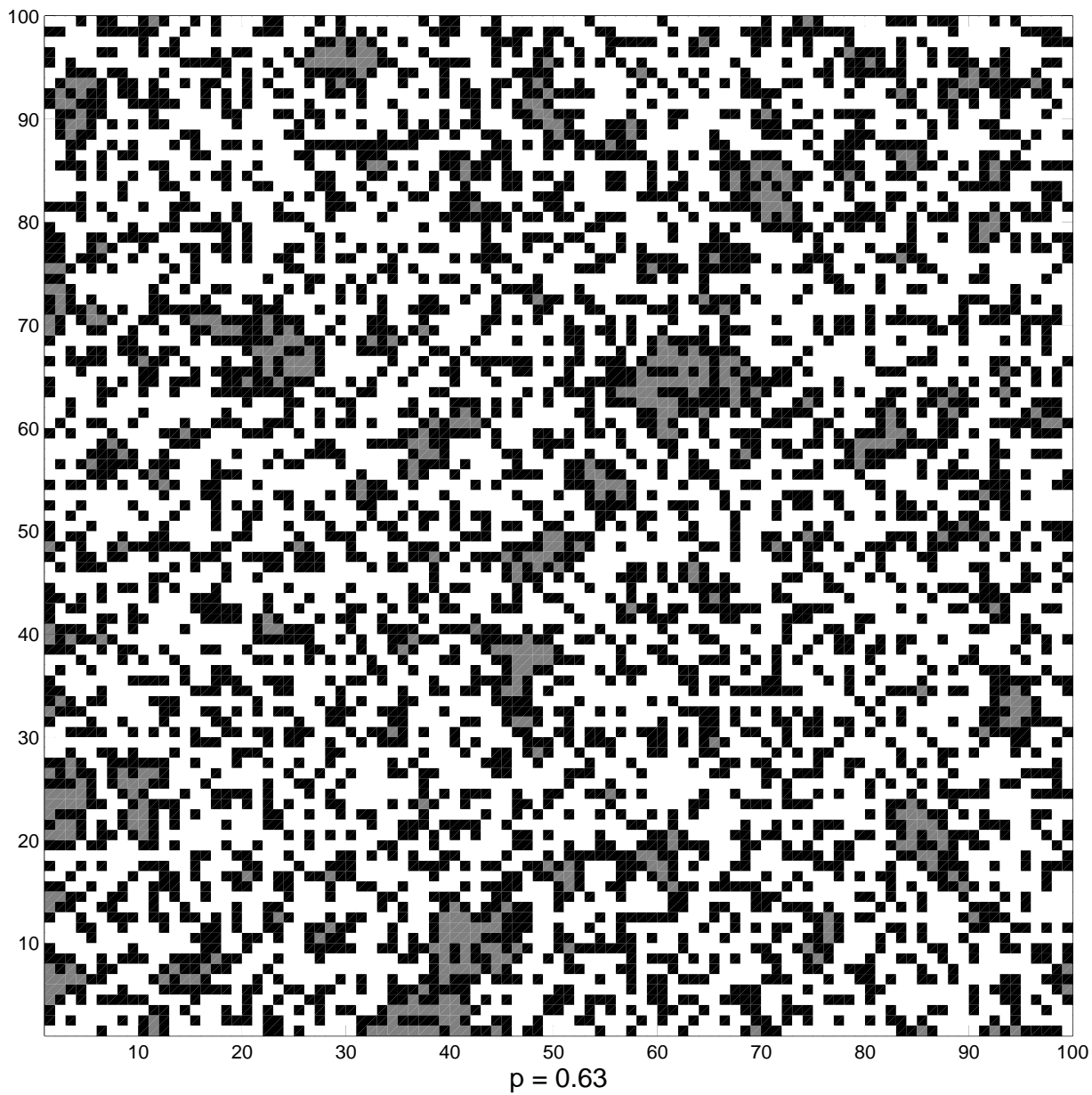}
\caption{Site percolation on a square lattice.   Concentration of
conducting pixels is equal  to $p=0,63 $. Grey cells of a square
lattice correspond to the conducting pixels isolated from maximal
(white) cluster}\label{f:pint2}
\end{figure}

At a low concentration $p$, the conductor sites are either
isolated or form small clusters of nearest neighbor sites (see
figure~\ref{f:pint1}). We suppose that two conductor sites belong
to the same cluster if they are connected by a path of nearest
neighbor conductor sites, and a current can flow between them. At
low $p$ values, the mixture is an insulators, since a conducting
path connecting opposite edges of our lattice does not exist. At
large $p$ values, on the other hand, many conducting paths between
opposite edges exist, where electrical current can flow, and the
mixture is a conductor (see figure~\ref{f:pint2}).

\begin{figure}[h!]
\centering\noindent\vspace{1cm}
\includegraphics[width=.8\textwidth]{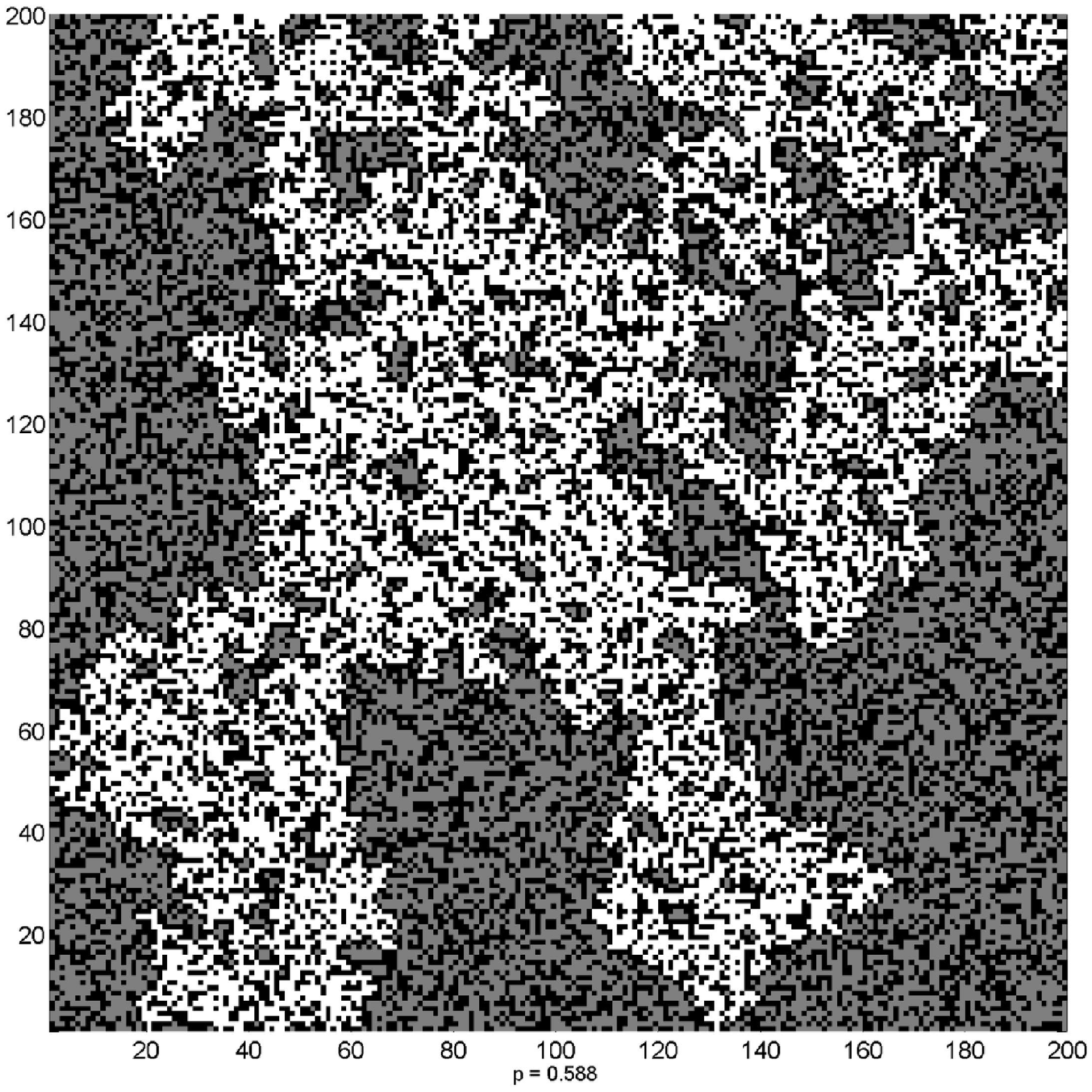}
\caption{Site percolation on a square lattice.  Concentration of
conducting pixels is equal  to $p=0,588$} \label{f:pint3}
\end{figure}

At some concentration in between, therefore, a threshold
concentration $p_{c}$ must exist where for the first time
electrical current can percolate from one edge to the other (see
figure~\ref{f:pint3}). Thus, for the values $p < p_{c}$ we have an
insulator, and for $p \ge p_{c}$ we have a conductor. The
threshold concentration is called the \textit{percolation
threshold}, or, since it separates two different phases, the
\textit{critical concentration}. For a site problem on  a square
lattice the percolation threshold is approximately equal to 0.59,
i.e., $p\approx0.59 $ \cite{BundeHav1,BundeHav,Feder,Stauff}. A situation for a value $p$ close to the
threshold is displayed in Figure~\ref{f:pint3}.

If the occupied sites are superconductors and the empty sites are
conductors, then $p_{c}$ separates a normal-conducting phase for
values $p<p_{c}$ transition from a superconducting phase where $p
\ge  p_{c}$. Another example is a mixture of magnets and
paramagnets, where the system changes at $p_{c}$ from a paramagnet
to a magnet.

In contrast to the more common thermal phase transitions, where
the transition between two phases occurs at a critical
temperature, the percolation transition described here is a
geometrical phase transition, which is characterized by the
geometric features of large clusters in the neighborhood of
$p_{c}$. At low values of $p$ only small clusters of occupied
sites exist. When the concentration $p$  increases, the average
size of the clusters increases, as well. At the critical
concentration $p_{c}$, a large cluster appears which connects
opposite edges of the lattice.
This cluster commonly named
\textit{spanning cluster} or \textit{percolating cluster} \cite{BundeHav1,BundeHav,Feder,Stauff}. In the thermodynamic
limit, i.e. in the infinite system limit spanning cluster named \textit{infinite
cluster}, since its size diverges when the size
of the lattice   increases to infinity.
It should be emphasized here that from traditional standpoint there exist
unique \textit{infinite} cluster and this \textit{infinite} cluster
always coincides with \textit{spanning} cluster \cite{BundeHav1,BundeHav,Feder,Stauff}.

When $p$   increases
further, the density of the infinite cluster also increases, since
more and more sites  start to be a part of the infinite cluster.
Simultaneously, the average size of the finite clusters, which do
not belong to the infinite cluster, decreases. At $p=1$,
trivially, all sites belong to the infinite cluster.
In percolation, the concentration $p$ of occupied sites plays the
same role as the temperature in thermal  phase transitions.
Similar to thermal transitions, long range correlations control
the percolation transition and the relevant quantities near
$p_{c}$ are described by power laws and critical exponents \cite{BundeHav1,BundeHav,Feder,Stauff}.

The percolation transition is characterized by the geometrical
properties of    clusters for values of $p$ that are close to
$p_{c}$. One of  important characteristics describing these
properties  is the probability, $ P_{\infty }$, that a site
belongs to the infinite cluster. For $ p<p_{c}$, only finite
clusters exist, and, therefore, it follows $P_{\infty }=0$. For
values $p>p_{c}$, $P_{\infty }$ behaves similarly to the
magnetization below critical temperature, and increases
with $p$ by a power law
\begin{equation}\label{perc1}
P_{\infty }\sim \left( p-p_{c}\right) ^{\beta },
\end{equation}
where $\beta=5/36$ is critical exponent in 2D case \cite{BundeHav1,BundeHav,Feder,Stauff}.

The linear size of the finite clusters, below and above  percolation transition, is characterized by
the $correlation$ $length$ $\xi $. The correlation length is defined as the
mean distance between two sites on the same finite cluster. When $p$
approaches $p_{c}$, $\xi $ increases as
\begin{equation}\label{perc2}
\xi \simeq a\cdot\left| p-p_{c}\right| ^{-\nu },
\end{equation}
with the same exponent $\nu=4/3$ below and above the threshold \cite{BundeHav1,BundeHav,Feder,Stauff}.

To obtain $\xi $ 
averages over all finite clusters in the lattice are required.

That is whelsy to note that all quantities described above are defined
in the thermodynamic limit of large systems. In a finite system,
$P_{\infty }$, for example, is not strictly zero below $p_{c}.$ 

The structure of percolation cluster can be well described in the
framework of the fractal theory. We begin by considering the
percolation cluster at the critical concentration $p_{c}$. A
representative example of the \textit{spanning} clusters shown in Fig.
\ref{f:pint3}. As seen in the figure, the \textit{infinite} cluster
contains holes of all sizes. The cluster is self-similar on all
length scales (larger than the unit size and smaller than the
lattice size), and can be regarded as a fractal. The
\textit{fractal dimension}, $d_{f}$, describes how, on the
average, the mass, $M$, of the cluster
within a sphere of radius $r$ scales with the $r,$%
\begin{equation}\label{tag4.1}
M\left( r\right) \sim r^{d_{f}}.
\end{equation}
In random fractals, $M\left( r\right) $ represents an average over
many different cluster configurations or, equivalently, over many
different centers of spheres on the same \textit{infinite} cluster. Below
and above $p_{c}$, the mean size of the finite clusters in the
system is described by the correlation length $\xi $. At $p_{c}$,
$\xi $ diverges and
holes occur in the \textit{infinite} cluster on all length scales. Above $p_{c}$, $%
\xi $ also represents the linear size of the holes in the \textit{infinite} cluster.
Since $\xi $ is finite above $p_{c}$, the \textit{infinite} cluster can be
self-similar only on length scales smaller than $\xi $. We can interpret $%
\xi \left( p\right) $ as a typical length up to which the cluster is
self-similar and can be regarded as a fractal. For length scales larger than
$\xi $, the structure is not self-similar and can be regarded as
homogeneous. If our length scales is smaller than $\xi $, we see a fractal
structure. On length scales larger than $\xi $, we see a homogeneous system
which is composed of many unit cells of size $\xi $. Mathematically, this
can be summarized as
\begin{equation}\label{per0101}
M\left( r\right) \sim \left\{
\begin{array}{ll}
r^{d_{f}}, & r\ll \xi , \\
r^{d}, & r\gg \xi .
\end{array}
\right.
\end{equation}

One can relate the fractal dimension $d_{f}$ of percolation cluster to the
exponents $\beta $ and $\nu $ \cite{BundeHav1,BundeHav,Feder,Stauff}. The probability that an arbitrary site within
a circle of radius $r$ smaller than $\xi $ belongs to the \textit{infinite} cluster,
is the ratio between the number of sites on the \textit{infinite} cluster and the
total number of sites,
\begin{equation}\label{tag4.3}
P_{\infty }\sim \frac{r^{d_{f}}}{r^{2}},  \hspace{5mm} r<\xi .
\end{equation}
This equation is certainly correct for $r=\lambda \xi $, where $\lambda $
is an arbitrary constant smaller than 1. Substituting $r=\lambda $ $\xi $ in
(\ref{tag4.3}) yields
\begin{equation}\label{tag4.4}
P_{\infty }\sim \lambda^{d_f-2}\cdot\frac{\xi ^{d_{f}}}{\xi ^{2}}\sim\frac{\xi ^{d_{f}}}{\xi ^{2}}.
\end{equation}
Both sides are powers of $p-p_{c}$. By substituting
(\ref{perc1}) and (\ref{perc2}) into (\ref{tag4.4}) we obtain,
\begin{equation}\label{tag4.5}
d_{f}=2-\frac{\beta }{\nu }.
\end{equation}
Thus the fractal dimension of the \textit{infinite} cluster at $p_{c}$ is not a new
independent exponent but depends on $\beta $ and $\nu $. Since $\beta $ and $%
\nu $ are universal exponents, $d_{f}$ is also universal. It can
be shown \cite{Stauff} that
(\ref{tag4.5}) also represents the fractal dimension of the finite
clusters at $p_{c}$ and below $p_{c}$, as long as their linear
size is smaller than $\xi $.

The exponents $\beta $, $\nu $, and $\gamma $ describe the critical behavior
of typical quantities associated with the percolation transition, and are
called the $critical$ $exponents$. The exponents are universal and depend
neither on the structural details of the lattice (e.g., square or
triangular) nor on the type of percolation (site, bond, or continuum), but
only on the dimension $d$ of the lattice ($d=2$ in our present consideration).

This universality is a general feature of phase transitions, where
the order parameter vanishes continuously at the critical point
(second order phase transition). In Table 1, the values of the
critical exponents $\beta $, $\nu $, and $ \gamma $ in percolation
are listed for 2D case \cite{BundeHav1}.

\vspace{0.2in}
\begin{flushright}
Table 1.
\end{flushright}
\begin{center}
\begin{tabular}{|l|c|}\hline
\bf{Percolation} & $d=2$ \\ \hline
$Order$ $parameter$ $P_{\infty }:\beta $ & 5/36   \\ \hline
$Correlation$ $length$ $\xi :\nu $ & 4/3  \\ \hline
$Mean$ $cluster$ $size$ $S:\gamma $ & 43/18       \\
\hline
$Fractal$ $dimention$ & $91/48$ \\ \hline
\end{tabular}
\end{center}
\vspace{0.2in}

The fractal dimension, however, is not sufficient to fully
characterize a percolation cluster. For a further intrinsic
characterization of a fractal we consider the shortest path
between two sites on the cluster. We denote the length of this
path, which is called the `chemical distance', by $l$ \cite{BundeHav1,BundeHav,Feder,Stauff}. The $graph$
dimension $d_{l}$, which is also called the `chemical' or
`topological' dimension, describes how the cluster mass $M$ within
the chemical distance $l$ from a given site scales with $l$,
\begin{equation}\label{tag4.6}
M\left( l\right) \sim l^{d_{l}}.
\end{equation}
While the fractal dimension $d_{f}$ characterizes how the mass of
the cluster scales with the  Euclidean  distance $r$, the graph
dimension $d_{l}$ characterizes how the mass scales with the
chemical distance $l$.

The concept of the chemical distance also plays an important role in the
description of spreading phenomena such as epidemics and forest fires,
which propagate along the shortest path from the seed.

Let us investigate the percolation problem from positions of the
new arithmetics of infinite and infinitesimal numbers (see
\cite{Sergeyev,chaos,informatica}). Consider 
a 2D square
lattice with period $a $ and the linear size
$L=a\cdot\mbox{\ding{172}}$. The full number of cells of such a
lattice   is, therefore, infinite and is equal to $V
=\mbox{\ding{172}}^2 $. Since the critical parameter is defined as the
attitude of the occupied sites number $N$ to their full number $p=N/V=N/\mbox{\ding{172}}^2$ then the
smallest change in concentration $\delta p=\mbox{\ding{172}}^{-2}$ is equivalent to adding or substracting
only one occupied site.
The infinitesimal small value $\delta p$ is the maximum precision level we can distinguish by concidering the critical parameter $p$ on the $\mbox{\ding{172}}\times\mbox{\ding{172}}$
lattice.
In order to obtain a higher precision level we should increase our lattice
linear size. For example, if we use a lattice with period $a$ and linear size
$L=a\cdot\mbox{\ding{172}}^{1+\vartheta/2}$, where $\vartheta>0$, the maximum precision level we can distinguish by considering the critical parameter $p$ is
$\delta p=1/V=\mbox{\ding{172}}^{-(2+\vartheta)}$.

When we investigate the percolation problem we increase or decrease the critical parameter $p$ using an appropriate precision level $\delta p$ starting from an arbitrary point in between $p=0$ and $p=1$.
According to the \textbf{Postulate 1} we are able to
execute only a finite number of steps with length $\delta p$. Therefore,
the length of critical parameter $p$ interval that we can investigate is determined
by the precision level we chose.

\begin{figure}[h!]
\centering\noindent\vspace{1cm}
\includegraphics[width=.9\textwidth]{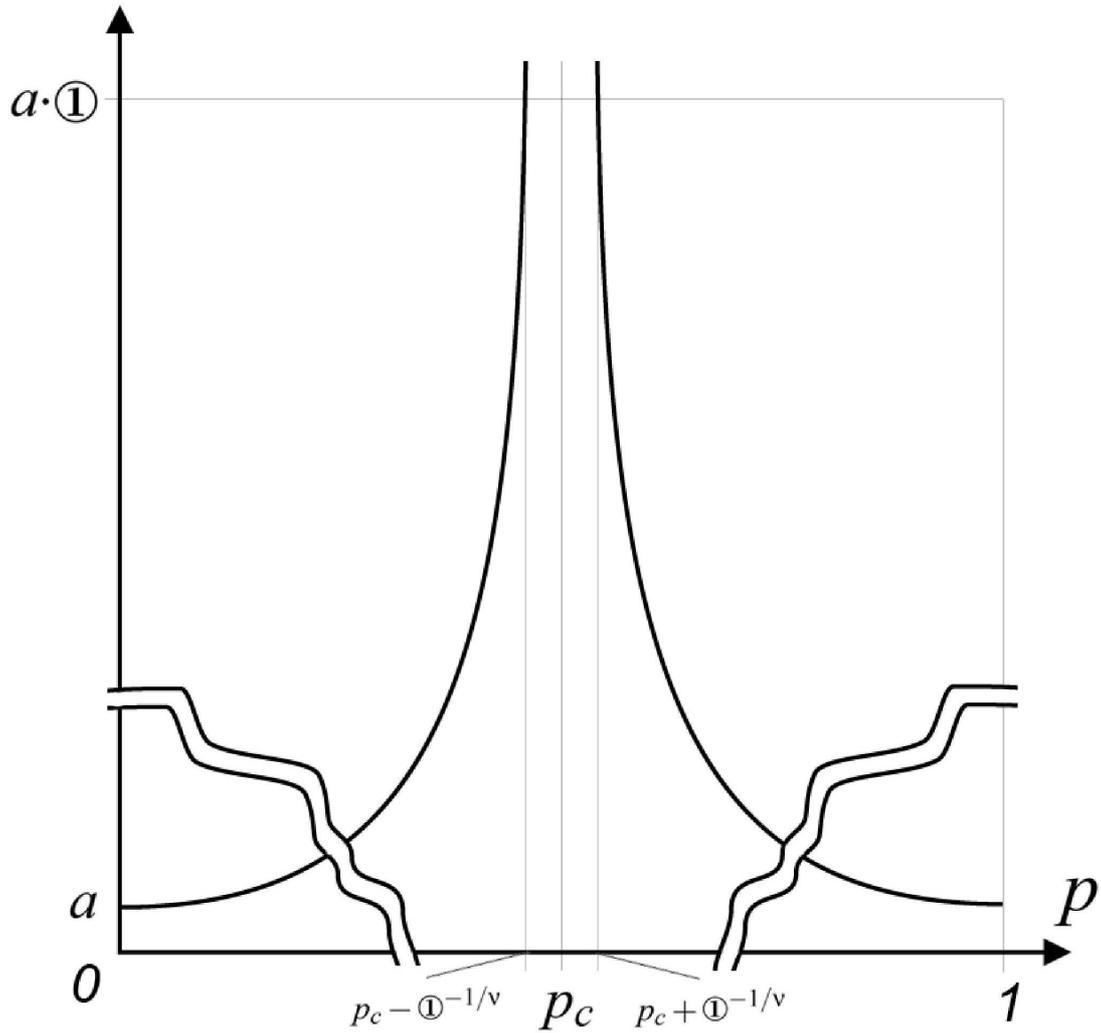}
\caption{Correlation length versus $p$}
\label{perinf:1}
\end{figure}

Concider the behavior
of correlation radius. In the vicinity of percolation
threshold the correlation radius diverges according to (\ref{perc2}).
On the other hand, the radius of correlation cannot exceed the system
linear size $\xi\lesssim \xi_{max}=L=a\cdot\mbox{\ding{172}}$,
where $\xi_{max}=a\cdot\mbox{\ding{172}}$ is the maximal correlation length.
The situation is depicted in Figure~\ref{perinf:1}.
We see that in the range $[p_c-\mbox{\ding{172}}^ {-1/\nu},\;p_c+\mbox{\ding{172}}^{-1/\nu}]$
the radius of correlation in our $\mbox{\ding{172}}\times\mbox{\ding{172}}$ lattice does not change and keeps the value $\xi_{max}=a\cdot\mbox{\ding{172}}$.

Now we should decide which step we shall use to express different
points on $p$ axis. Infinitely many variants can be chosen dependent on the
precision level we want to obtain. All these variants form three groups.
The first group appears when in order to change $p$ we use a small but still finite step $\delta p\ll1$. In the case the phase transition is infinitely
sharp because $\delta p\gg\mbox{\ding{172}}^ {-1/\nu}$. The second group appears when $\delta p=c\cdot\mbox{\ding{172}}^{-1/\nu}$,
where $c$ is a finite grossdigit that is less than one. In the case the phase transition occupies the finite interval $[p_c-\mbox{\ding{172}}^ {-1/\nu},\;p_c+\mbox{\ding{172}}^{-1/\nu}]$.
The third group appears when $\delta p=\mbox{\ding{172}}^{-\varsigma}$,
where $\displaystyle\frac{1+\nu}{\nu}\leq\varsigma\leq2$.
\footnote{
For example, if we add only one occupied site
in our greed, then $p$ increases by $\delta p=\mbox{\ding{172}}^{-2}$,
and that is the smallest step along $p$ we can distinguish in our $\mbox{\ding{172}}\times\mbox{\ding{172}}$ lattice.} In the case phase transition interval contains more than $\mbox{\ding{172}}$
different points and if we execute a finite number of steps with length $\delta p$ along this infinite transition area there exist three possibilities: 1) system contains a lot of finite
and \textit{infinite}
clusters that coagulate but \textit{spanning} cluster is still absent; 2) \textit{spanning}
cluster already exists and absorbs finite
and infinite
clusters; 3) at the beginning of our execution \textit{spanning} cluster is absent
but it appears after finite number of steps.
This appearance is due to adding
only one occupied site in our grid that produce confluence of either two infinite clusters or one finite
and one infinite clusters.

\begin{figure}[h]
\centering\noindent\vspace{1cm}
\includegraphics[width=.8\textwidth]{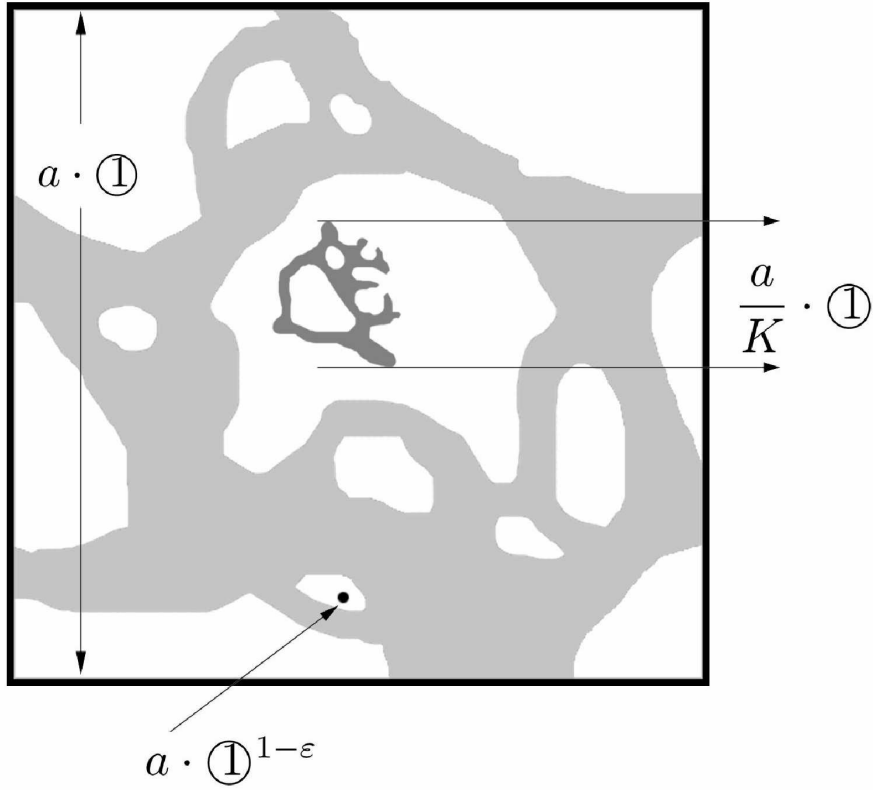}
\caption{Embedded infinite clusters of different scales}
\label{ff:21}
\end{figure}

  Figure  \ref{ff:21} shows that \textit{spanning}
cluster could envelop a set of embedded infinite clusters of
different scales when we choose the critical parameter
infinitesimally close to percolation threshold value.
  One infinite cluster is embedded into another also
infinite but already spanning cluster. This situation is similar
to that with finite clusters, when in finite system one finite
cluster is embedded into another also finite but already spanning
cluster.  Some of the embedded infinite clusters are comparable
with the \textit{spanning} cluster and have linear sizes $R$ that
could be expressed by following:
\begin{equation}
R=\displaystyle\frac{a}{K}\mbox{\ding{172}},
\end{equation}
where $K>1$ is a finite number. Remainder of the embedded infinite clusters has linear sizes that are indefinitely small as compared with $\mbox{\ding{172}}$ and could be expressed as $R=\varepsilon\cdot\mbox{\ding{172}}$, where $\varepsilon$ is infinitesimal number.

\begin{figure}[h!]
\centering\noindent\vspace{1cm}
\includegraphics[width=\textwidth]{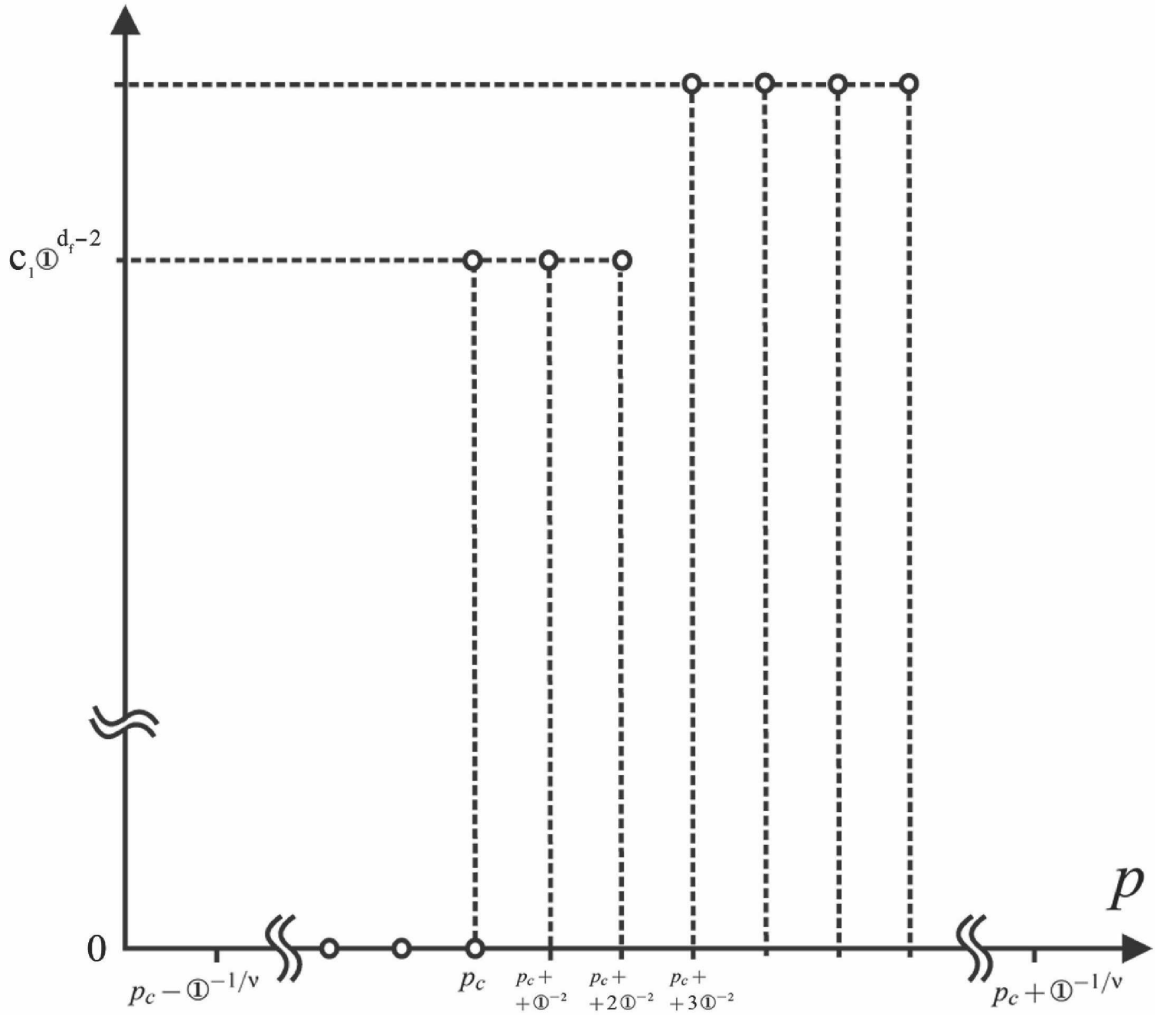}
\caption{${P_\infty}$ vs $p$}
\label{Pinf}
\end{figure}

On the step when \textit{spanning} cluster appears the order parameter jumps from
zero value up to the infinitesemal value as seen in Fig.~\ref{Pinf}
\begin{equation}
{P_\infty}_{min}\sim C_1\mbox{\ding{172}}^{d_f-2}=C_1\mbox{\ding{172}}^{-\beta/\nu}=
C_1\mbox{\ding{172}}^{-\frac{5}{48}},
\label{perc011}
\end{equation}
where $C_1$ is a finite number. The first equality in
(\ref{perc011}) defines ${P_\infty}_{min}$ as a measure of the
relative size of \textit{spanning} cluster expressible as a
proportion of elements number of spanning cluster
$C_1\mbox{\ding{172}}^{d_f}$ to total number of grid elements
$C_1\mbox{\ding{172}}^{2}$. The second equality in (\ref{perc011})
appears as a consequence of Eq.~(\ref{tag4.5}). Actually, spanning
clusters could come in different size and shapes, so the constant
$C1$ depicted in Fig. 6 could vary distinctly. In addition, when
spanning cluster already exists and we increase $p$ by adding new
occupied sites the spanning cluster could expand because it
engrosses other finite and infinite clusters.

We see that
application of the new arithmetic of infinite  and infinitesimal
numbers gives us a unique opportunity to consider a point of phase
transition in more detail (viewed just like a point with respect of
traditional approach).


\section{Gradient percolation}\label{s4}

\begin{figure}[h!]
\centering\noindent\vspace{1cm}
\includegraphics[width=.75\textwidth,height=12cm]{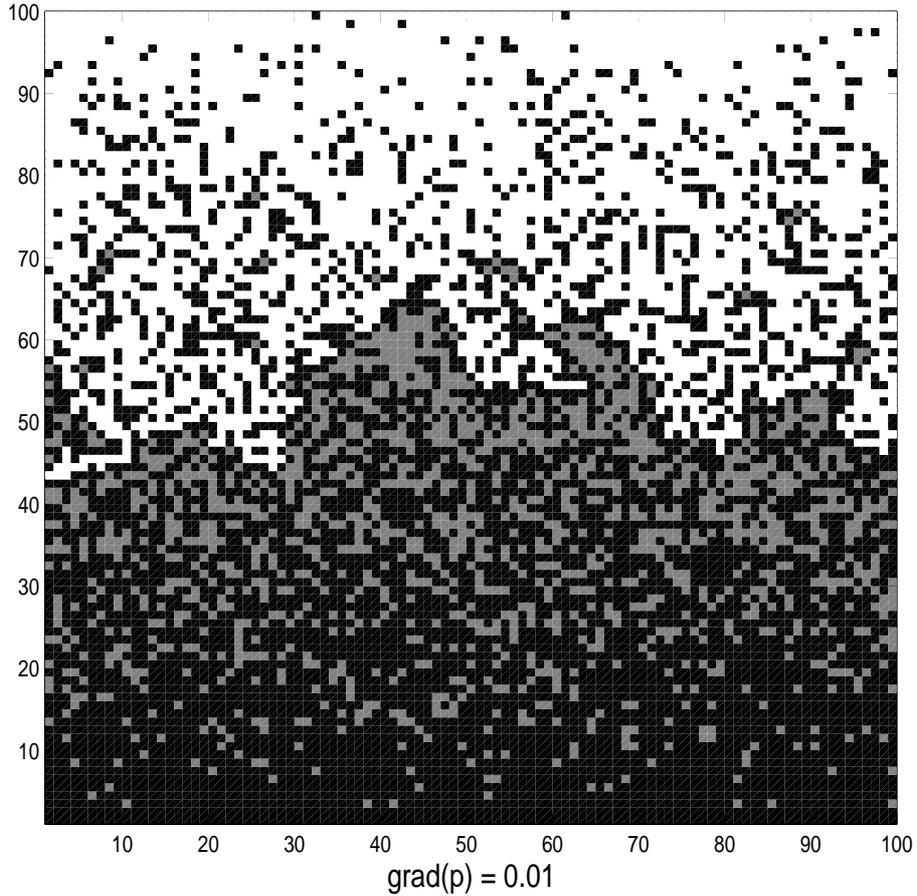}
\caption{Gradient percolation in two dimensions} \label{grad}
\end{figure}

An important site-percolation problem generalization appears when
the concentration $p$ of occupied sites varies with the vertical
distance $z$ in our square grid. In literature (see
\cite{Gouyet:94}), this
generalization is commonly named as the \textit{gradient
percolation}. It can be conveniently pictured in a geographical
description in which the set of sites connected to the area
$p\lesssim1$ is called the `land'.  In Fig~\ref{grad}, it is shown
by white pixels. In this geographical language the set of
connected empty sites not surrounded by land is called the `sea',
in Fig~\ref{grad}, it is shown by black pixels. Then, there
naturally exist groups of occupied sites that are not connected
with the land called `islands'. They are  shown by grey pixels in
Fig~\ref{grad}. Analogously, there exist also connected empty
sites surrounded by the  land. They are calle `lakes', which are
shown also in black in   Fig~\ref{grad}. In this geographical
description, the part of the land in contact with the sea is
called the `seashore'. In \cite{Gouyet:94} this line is attributed
as the \textit{diffusion front}.

The diffusion front is conveniently described (see
\cite{Gouyet:94}) by its average
width  $h_f$, that can be related easily to the concentration
gradient $dp/dz$  at the position of the front. We see in
Fig.~\ref{grad} that, far from the front, islands or lakes are
very small, whereas, near the front, their size becomes comparable
to the width of the front. The islands correspond to the finite
clusters in a percolation system, and the lakes correspond to the
finite holes. The typical linear size of both quantities scales as
$\xi$. Relation (\ref{perc2}) tells that the size of the
islands or lakes should increase when approaching the mean
position of the front. But this size, even at $z_f$, is bounded
due to the finite gradient of $p(z)$.
The maximum typical size of islands and lakes is then given by the width
of the front, which represents the only characteristic length scale in the
problem, and we can assume
\begin{equation}\label{math/g1}
h_f\simeq \xi(z_c\pm h_f).
\end{equation}
This assumption expresses our observation that islands or lakes near the front have the size
comparable to the width of the front. Using
(\ref{math/g1}) and expanding $p(z)$ around $z=z_c$ we obtain
\[
h_f\simeq a|p(z_c\pm h_f)-p_c|^{-\nu}\simeq a\Big|h_f\frac{dp}{dz}(z_c)\Big|^{-\nu},
\]
which gives 
\begin{equation}\label{math/g3}
h_f\simeq a^{\frac{\beta_f}{\nu}}\Big|\frac{d p}{dz}(z_c)\Big|^{-\beta_f},
\end{equation}
where
\begin{equation}\label{math/g4}
\beta_f=\displaystyle\frac{\nu}{1+\nu}.
\end{equation}
As percolation is a critical phenomenon, the exponent $\beta_f$
depends only on the dimensionality of the system (for $d=2$ it
follows $\beta_f=4/7$), and not on the particular lattice
structure (square, triangular, etc.).

Let us assume now, that we examine the gradient percolation
phenomenon on a square lattice $N^2 $ where $N =
\mbox{\ding{172}}$, and the critical parameter $p $ changes
linearly, accepting infinitesimal value $p(z=a)=\frac {1} {a}
\cdot\mbox{\ding{172}}^ {-1}$ (value equal to zero) in the first
line of lattice cells and value equal to unit
$p(z=a\cdot\mbox{\ding{172}})=1$ in the last, $
\mbox{\ding{172}}$-s. In other words,
\begin{displaymath}
p(z)=A\cdot z,
\end{displaymath}
where $A=\frac {1} {a} \cdot\mbox{\ding{172}}^ {-1}$ and $z$ changes discretely.
Then the value of the derivative in (\ref
{math/g3}) is $ \frac {1} {a} \cdot\mbox{\ding{172}}^ {-1} $, and the
diffusion front width makes
\begin{equation}\label{math/g31} h_f\simeq
a\cdot\mbox{\ding{172}}^{\beta_f}=a\cdot\mbox{\ding{172}}^{4/7},
\end{equation}

Thus,  on scales of the observation commensurable with the size of
the entire system, the diffusion front width is viewed as
infinitesimally small and it is represented by the sharp border of
two contrast phases -- `sea' and `land'. On the contrary, length
scales commensurable with the finite number of the lattice periods
$a$ are completely absorbed by huge fluctuations of the front.
At last, on scales
proportional with $h_f$ the width of front appearers to the
observer as a finite value. 

\section{A brief conclusion}

In this paper, it has been shown that infinite and infinitesimal
numbers introduced in \cite{Sergeyev,informatica,Lagrange} allow
us to obtain exact numerical results instead of traditional
asymptotic forms at different points at infinity. We consider a
number of traditional models related to the percolation theory
using the new computational methodology. It has been shown that
the new computational tools allow one to create new, more precise
models of percolation and to study the existing models more in
detail. The introduction in these models new, computationally
manageable notions of the infinity and infinitesimals gives a
possibility to pass from the traditional qualitative analysis of
the situations related to these values to the quantitative one.
Naturally, such a transition is very important from both
theoretical and practical viewpoints.

The point of view presented in this paper  uses
strongly two methodological ideas borrowed from Physics:
relativity and interrelations holding between the object of an
observation and the tool used for this observation. The latter is
directly related to connections between Analysis and Numerical
Analysis because the numeral systems we use to write down numbers,
functions, etc. are among our tools of investigation and, as a
result, they strongly influence our capabilities to study
mathematical objects.

Note that  foundations of Analysis have been developed more than
200 years ago with the goal to develop mathematical tools allowing
one to solve problems arising in the real world, as a result, they
reflect  ideas that people had about Physics in that time. Thus,
Analysis that we use now does not include numerous achievements of
Physics of the XX-th century. The brilliant efforts of Robinson
made in the middle of the XX-th century have been also directed to
a reformulation of the classical Analysis in terms of
infinitesimals and not to the creation of a new kind of Analysis
that would incorporate new achievements of Physics. In fact, he
wrote in paragraph 1.1 of his famous book \cite{Robinson}: `It is
shown in this book that Leibnitz' ideas can be fully vindicated
and that they lead to a novel and fruitful approach to classical
Analysis and to many other branches of mathematics'.

Site percolation and gradient percolation have been studied by
applying the new computational tools. It has been established that
in infinite system phase transition point is not really a point as
with respect of traditional approach. In light of new arithmetic
it appears as a critical interval, rather than a critical point.
Depending on ``microscope" we use this interval could be regarded
as finite, infinite and infinitesimal short interval. Using new
approach we observed that in vicinity of percolation threshold we
have {\it many} different {\it infinite clusters} instead of {\it
one infinite cluster} that appears in traditional consideration.
Moreover, we have now a tool to distinguish those infinite
clusters. In particular, we can distinguish {\it spanning
infinite} clusters from \textit{embedded} infinite clusters.

Than we consider gradient percolation phenomenon on infinite
square lattice with infinitesimal gradient of critical parameter
$p$ that changes linearly, accepting infinitesimal value
$p(z=a)=\frac {1} {a} \cdot\mbox{\ding{172}}^ {-1}$ (value equal
to zero) in the first line of lattice cells and value equal to
unit $p(z=a\cdot\mbox{\ding{172}})=1$ in the last, $
\mbox{\ding{172}}$-s line of lattice cells. We observe that
diffusion front width in this case stretches for an infinite
number of lattice spacing: $h_f\simeq
a\cdot\mbox{\ding{172}}^{\beta_f}=a\cdot\mbox{\ding{172}}^{4/7}$.
And again this value could be regarded as finite, infinite and
infinitesimal short depending on ``microscope" we use.

\section*{Acknowledgements}
This work is partly supported by the Russian Foundation for Basic
Research (project No. 10-01-00690-a, 11-05-12055-ofi-m-2011,
11-04-12144-ofi-m-2011), by the Ministry of education and science
of Russian Federation (project No. 2.1.1/6020), by Russian Federal
Programs (Nos. 16.740.11.0488, 14.740.11.0075, 16.512.11.2136),
supported by the Russian Foundation for Basic Research (project
No. 10-05-01045-a), jointly by the Russian Foundation for Basic
Research and the Government of the Nizhny Novgorod region (Nos.
11-01-97028-``volga", 11-04-97015-``volga", 11-06-97024-``volga",
11-02-97046-``volga", 11-05-97037-``volga", 11-02-97101-``volga"),
by Russian Federal Program (No. 16.740.11.0488) and the Ministry
of education and science of Russian Federation (project No.
2.1.1/6020).




\end{document}